\magnification=\magstep1

\hsize=6truein
\vsize=8.5truein
\hfuzz 6pt
\null 
\tenrm 
\def \ir{{\rm i}}
\def \e{{\rm e}}
\def \um{{\textstyle {1\over2}}}

{\centerline
{VERY STRONG AND SLOWLY VARYING MAGNETIC FIELD}}\vskip 1pc
{\centerline
{ AS SOURCE OF AXIONS}}

\vskip 2pc
{\centerline
{ Giorgio CALUCCI\footnote*{E-mail: giorgio@ts.infn.it}}}
\vskip 1pc
{\centerline
{\it Dipartimento di Fisica Teorica dell'Universit\`a di Trieste,
Trieste, I 34014 Italy}}
{\centerline
{\it INFN, Sezione di Trieste, Italy}}
\vskip 2pc
{\centerline {Abstract}}
\narrower
\midinsert
{The investigation on the production of particles in slowly varying but extremely intense magnetic field in extended to the case of axions. The motivation is, as for some previously considered cases, the possibility that such kind of magnetic field may exist around very compact astrophysical objects.}
\endinsert
\vfill
 \eject
\vskip 1pc 
{\bf 1. Statement of the problem}
\vskip 1pc  
A magnetic field of huge strength can give rise to real particles even if its rate of variation is very small: this posibility could be of some interest from a pure theoretical point of view, but it gains more physical relevance if one accepts that such kind of field configurations may be present around some very compact astrophysical objects[1-3]. In this case the time variation is related to the  evolution of the source, by collapse, rotations or else, and it is therefore very slow, in comparison with the times typical of elementary-particle processes. We can call the former time the macroscopic time and the latter the microscopic one.
The production of light particles in these processes has been analyzed in some detail in some previous papers[4],with the suggestion that it is one of the mechanisms at work in the phenomenon of {\it gamma-ray bursts}[5]; the typical microscopic time is related to the electron mass since photons are produced through real or virtual intermediate states of $e^--e^+$-pairs. The  lightest particles that could be produced are massive neutrinos, but the magnetic-moment coupling induced by the standard electroweak interactions is extremely small.\par
There is, at least in the theoretical realm, another very light particle, that is the axion[6]: owing to its dynamical characteristics it must be coupled also to the electromagnetic field[7], even more its electomagnetic coupling is being actively studied from an experimental side[8] and the possibility of detecting such particles as coming from nonterrestrial sources has already been foreseen[9]. It is immediately seen that the production of axions by a varying magnetic field must be realized through a mechanism different from the previously considered one, in fact the axions are coupled[7] only to the pseudoscalar density ${\bf E\cdot B}$ so the presence of an electric field is necessary as a starting point, but a nonstatic magnetic field creates always an electric field and even though the rate of variation is small, the very large magnetic strength makes the electric field not a tiny one. \par
In the present paper the coupling of the axions field with a given ${\bf E\cdot B}$ density is written in standard second-quantized formalism, then the effect of time variation of that density on the axion vacuum is determined and the consequent production is calculated. The result depends both on the spatial shape and on the time variation of the magnetic field: in accordance with the prevailing astrophysical hypotheses[1-3] the magnetic field is seen as a bundle of lines of force which may safely be considered straight in comparison with the microscopic scale. The time-variation could affect both the shape and the strength of the fields, both are effective in the production process. The  calculation procedure is not the standard adiabatic approximation [10] as used in previous investigations [4], but the
feature that one has to deal with a two-scale problem is still fully relevant.

\vskip 1pc  
 {\bf 2. General form of the production probability }
\vskip 1pc  
 The starting point is a second-quantized axion field in presence of given, classical, magnetic and electric fields. The axion field $\phi(x)$ is coupled to the pseudoscalar density \footnote*{the same kind of coupling is possible also for the neutral pion, but in this case other channels of production are present [11]} $G(x)={\bf E}(x)\cdot{\bf B}(x)$ and the coupling constant, of dimension of length, is here indicated by $C$. We assume that the interaction lasts from an initial time $t_o$ until a final time $t$.\par
 So we have for the axion field the expression:
  $$\phi(x)=\phi_o(x)+\chi(x)=\phi_o(x)+C\int\Delta_R(x-y)G(y)d^4y  \eqno (1)$$
  Here $\Delta_R(x-y)$ is the standard retarded Green function, the source is a $c-$number, so the same holds for $\chi$.
  The field $\phi$ is free before $t_o$, where it has the standard expansion:
  $$\phi_o(x)={1\over {(2\pi)^{3/2}}}\int{d^3k\over{\sqrt{2\omega_k}}}
  \Big[ a_o({\bf k})\e^{\ir{\bf k\cdot r}-\ir\omega t}+ 
   a_o^{\dagger}({\bf k})\e^{-\ir{\bf k\cdot r}+\ir\omega t}\Big]\,, \eqno(2)$$ then it acquires a contribution from $\chi$, this term has the following actual expression:
  $$\chi(x)={\ir\over {(2\pi)^{3/2}}}\int{d^3k\over{2\omega_k}}\Big[\e^{-\ir\omega_k t}\int_{t_o}^t\!
  \e^{\ir\omega_k \tau}g_k(\tau) d\tau \e^{\ir {\bf k\cdot r}}-c.c.\Big] 
\eqno(3.1)$$
  $$g_k(t)={C\over {(2\pi)^{3/2}}}\int G({\bf r},t) \e^{-\ir {\bf k\cdot r}} \eqno(3.2)$$
  The reality condition for $G$, that gives $g^*_k=g_{-k}$ have been used,
  together with the initial condition $\dot\chi(t_o)=0$. The separation into
  positive and negative frequencies in $\chi(x)$ is unambiguous until the 
  typical frequencies of $G$ are small.
  
  \vskip 1pc 
 
  Having found the time evolution of the field the total production of axions is
  calculated in Heisenberg description of motion, $i.e.$ we take an initial state $a_o({\bf k})|\circ>=0\quad \forall {\bf k}$, $i.e.$ the vacuum in the absence of interaction then we express the mean particle number as the time dependent term $${\cal N}({\bf k},t)=<\circ | a^{\dagger}({\bf k},t)\; a({\bf k},t)|\circ>$$
  Since all the effect of the interaction is a $c-$number shift on the operators $a({\bf k},t)=a_o({\bf k})+b({\bf k})$ the calculation is easy, in particular when the interaction no longer acts we get 
  $${\cal N}_{f}({\bf k})=|b_f({\bf k})|^2 \eqno (4)$$
  The expression of $b_{f}({\bf k})$ can be read off from eq.s (2,3.1,3.2), it is:
  $$b_f({\bf k})={\ir\over{\sqrt{2\omega_k}}}\int_{t_o}^{t_f}\!
  \e^{\ir\omega_k \tau}g_k(\tau) d\tau \eqno (5)$$
  
  \vskip 1pc  
   
 {\bf 3. Detailed calculations }
\vskip 1pc  
   
 As anticipated in the introduction a more definite model of the magnetic field can by a field of uniform direction (at a given time) with some transverse shape, both the direction and the shape may vary in time, one possible restriction is the conservation of the total flux. More explicitly these conditions are realized by giving:
$${\bf A}={{F}\over{2\pi}}{{\bf n\wedge r}\over{r_{\perp}^2}}
      \big[1-w(\mu r_{\perp}) \big]\quad\quad   
 {\bf B}=-{{F}\over{\pi}r_{\perp}}{\bf n}\;\partial_{\perp}w(\mu r_{\perp})\;.
\eqno (6)$$
 The unit vector ${\bf n}$ gives the instantaneous direction of the magnetic field,
 $r_{\perp}=\sqrt{r^2-{\bf n\cdot r}^2}$ and $\partial_{\perp}$ the corresponding derivative, $F$ is the total magnetic flux; the parameter $\mu$ defines the size of the field in the transverse directions, $w$ is taken to be cylindrically symmetric and, obviously it must go to zero at infinity, the requirement $w(0)=1$ avoids singularities in $\bf E$.\par
Since ${\bf E}={-\bf\dot A}$ and  ${\bf A\cdot B}=0$ only the terms coming from the time variation of the direction of ${\bf A}$ contributes to
 $G= {\bf E\cdot B}$ so we get:
$$G=\Big({{F}\over {\pi}}\Big)^2{{\bf J\cdot r}\over {r^3_{\perp}}}
\big[(1-w)\partial_{\perp} w\big] \eqno (7)$$
Here ${\bf J}$ gives the angular velocity of ${\bf n}$ $i.e.\;{\bf J=n\wedge\dot n}$.
 In this configuration the Fourier transform of the source is 
$$ g_k(t)= 
{{\ir}\over {(2\pi)^{3/2}}}C(2F)^2\delta({\bf n\cdot k}) {\bf J\cdot k} S(k_{\perp}/\mu)\eqno (8.1)$$
$$S(k_{\perp}/\mu)=(\mu/k_{\perp})\int J_1(\rho k_{\perp}/\mu)\big[(1-w(\rho))w'(\rho)\big]d\rho /\rho\eqno (8.2)$$
Here $J_1$ is the Bessel function of order one and $w'$ indicates the derivative with respect to the argument. It is useful to remember that, owing to the presence of the factor $\delta({\bf n\cdot k})$ in the expression of $S$ we can substitute $k^2_{\perp}$ simply with $k^2$.\par

 We now remember that the model of magnetic field we have at hand is such that it is uniform along one direction, but this direction is continuously varying, so a most significant quantity is obtained by an angular integration
 $${\partial {\cal A}\over {\partial k}}=k^2\int d\Omega_k{\cal N}_{f} ({\bf k})\eqno (9)$$
 So we need a quantity like $\int d\Omega_k g_k(\tau)g_k^*(\tau')$ which contains a singularity due to the presence, in the domain of integration, of a $\delta$-square term which arises for $\tau=\tau'$. This is clearly due to the unphysical assumption that at every time there is a direction in which the pseudoscalar density $G$ is absolutely uniform. Since we are integrating over the direction at the end the effect of the singularity is mild enough, it results in a logarithmic divergence, however this fact must be explicitly death with, considering a finite extension of the fields. A more careful treatment is mathemetically heavier, so it is presented in an Appendix.\par
 The final result is given by
 $${\partial {\cal A}\over {\partial k}}\approx C^2 (2F)^4 {{k^2}\over{2\omega}}\,\Psi[S(k/\mu)]^2\;\sqrt{2}\ln 4kL\;.\eqno (10)$$
  \vskip 1pc  
If we give a definite transverse shape to the fields we get, evidently, a definite answer. In the situation where the fields change direction but not intensities, $i.e.$ we keep $\mu$ constant, we may give, tentatively the form $w(\rho)=\e^{-\rho^2}$. Then the expression for the function $ S(k/\mu)$ is[12]:
$$S(k/\mu)={{2\mu^2}\over {k^2}}\Big[\exp \Big(-{{k^2}\over{4\mu^2}}\Big)-
\exp \Big(-{{k^2}\over{8\mu^2}}\Big)\Big]\;.\eqno (11)$$
The limit $k\to 0$ of this expression is finite, so the whole production goes to zero only owing to the phase-space factor $k^2$ in front of the expression in eq.(10).
 This result, as it appears from the whole derivation, can be valid only for axion masses and so energies which are definitely larger than the typical frequencies of the astrophysical phenomena, no resonant dynamics is included.
 \vskip 1pc   
 {\bf 4. Some conclusions }
 \vskip 1pc  
The rate of production of axions by a slowly varying but very strong magnetic field has been calculated. The conditions are such that the astrophysical frequencies are very much lower than the proper frequencies of the axion field. In fact with an axion mass $m_A$ of the order of 1eV[13] the typical frequencies of the field shall exceed $10^{15}$ Hz and so no resonance conditions appear realistic. The relevant parameter turns out to be the dimension of the spatial inhomogeneity, in the present model $1/\mu$, here also is seems very reasonable to assume that $\mu<<m_A$. It is also possible to give an expression for the total number of produced particles.
$${\cal A}=\zeta {{C^2 F^4 \mu^3}\over {m_A} }\Psi\,\ln L\mu  \eqno(12)$$
Some of the factors owe their origin to the general form of the interaction, eq(1) and to dimensional requirament, it appers clear the role of the total rotation of the field $(\Psi)$ in determining the overall production; so when the rotation is uniform the rate is proportional to the angular velocity.\par
The numerical factor is more model dependent, in the chosen case it is 
$\zeta=8\sqrt {\pi} [2\sqrt {3}-2-\sqrt {2}]=0.704\dots$.\par
The transformation from the incoming Heisenberg field to the outgoing field can be implemented by the simple unitary operator 
$$ {\cal U}=\exp\Big[\int d^3k[ a({\bf k})b^*({\bf k})-a^{\dagger}({\bf k})b({\bf k})]\Big]$$ as:
$$ {\cal U} a({\bf k}){\cal U}^{\dagger}= a({\bf k})+b({\bf k})\;.$$
The actual form of the evolution operator gives the further information that the axion are produced with a Poissonian distribution of multiplicity, strictly speaking this is true for a production in a totally defined state, in operations like the one leading to eq. (9) this particular form can be blurred.
\vskip 1pc
{\it A very short mention on this problem was made at the XXXVI International Symposium on Multiparticle Dynamics, Paraty, R.J. - Brazil, September 2006}
 \vskip 1pc 
 {\bf Appendix }
 \vskip 1pc  
 We want to calculate $\int d\Omega_k g_k(\tau)g_k^*(\tau')$, with the functions $g$ given by eq.(8.1)
 and taking care of the finite extension of the magnetic field. This is implemented by substituting the $\delta-$functions as:
 $$\delta({\bf n\cdot k}) \to L \pi^{-1/2} \exp [-L^2({\bf n\cdot k})^2]\eqno (A.1)$$
  The calculation is performed in the particular case in which the rotation takes place in a constant plane, so ${\bf J}\|{\bf J'}$. The integration over the angles is performed in Cartesian coordinates, it is useful to introduce the unit vector of the three-momentum direction ${\bf k}=k{\bf v}$, then $\int d\Omega_k=2\delta (v^2-1)d^3v$ and through standard although lengthy calculations the representation is obtained:$$ \eqalign{&{\cal I}= \int d\Omega_k \delta({\bf n\cdot k}) {\bf J\cdot k}\delta({\bf n'\cdot k}) {\bf J'\cdot k} \to
  {1\over\pi}JJ'\times \cr  \int d\lambda &\exp[\ir\lambda(w^2-1)]w^2dw \big[({\bf n}\wedge{\bf n}')^2
  -2\ir\lambda/(Lk)^2-\lambda^2/(Lk)^4\big]^{-1/2} }\eqno (A.2) $$
  In the limit $L\to\infty$ it results
 $${\cal I}=JJ'/|{\bf n}\wedge{\bf n}'|$$
 which can be obtained in a simpler way.\par
Now we must integrate over $\tau$ and $\tau'$, times an oscillating factor $\e^{\ir\omega_k(\tau-\tau')}$. In the conditions that have been chosen the motion takes place in a plane, so $\bf n$ is characterized by a unique angle $\psi$ and $\bf n'$ by $\psi'$ hence the integration over time amounts at an angular integration, in fact $({\bf n\wedge n'})^2=(\sin (\psi-\psi'))^2$ and moreover $J=\dot\psi$ and
$J'=\dot\psi'$. So we must integrate $\cal I$ in $d\psi\,,\,d\psi'$ from 0 to some final angle $\Psi_f$. Defining $\gamma=\psi-\psi'$ we see that the integrand shows, in the limit $L\to\infty$, a singularity for $\gamma=0$, so we perform the integration from $-\um\pi$ to $\um\pi$ because the domain which do not include zero has no singular behavior, the oscillating factor is approximated with its value on the singular point $\tau=\tau'$, so that the exponential factor reduces to 1. Then the integral is a complete elliptic integral,[12] which can be conveniently expressed in term of the hypergeometric function.
In fact the integration in $d\gamma$ is:
$$X_{\gamma}=\int {{d\gamma}\over {\sqrt{\sin^2\gamma+Q}}} \quad {\rm with}\quad
Q=-{{2i\lambda}\over {(Lk)^2}}-{{\lambda^2}\over {(Lk)^4}} \eqno (A.3)$$
The result of the integration is
  $$X_{\gamma}=\um\pi\sqrt {{1\over {Q+1}}}\;{}_2F_1\Big(\um,\um;1;{1\over{Q+1}}\Big) $$
and in the limit $L\to\infty$, which gives $Q\to0$ we get;
$$X_{\gamma}\to \sqrt {2}\big[2\ln 2+\ln Lk\big]\;.$$
Since the dominant term in the limit is independent of the auxiliary parameter $\lambda$ the rest of the integration in eq. (A.2) is straightforward and it gives
 $${\cal I}=\Psi \sqrt{2} \ln 4kL \eqno (A.4)$$
{\it A comment:} what is excluded is the possibility that the magnetic field should perform more than a complete rotation, this would destroy the correspondence $\psi=\psi'\leftrightarrow \tau=\tau'$.

\vfill\eject

\vskip 1pc
 {\bf References}
\vskip 1pc 

\item {1.}C. Thompson, R.C. Duncan, Astrophys.J. 408 (1993) 194
\item {2.}H. Hanami, Astrophys.J. 491 (1997) 687
\item {3.}C. Kouveliotou, R.C. Duncan, C. Thompson, Sci.Am. 288,2 (2003) 34
\item {4.}G. Calucci, Mod.Phys.Lett. A14 (1999) 1183;\par
          A. DiPiazza, G.Calucci, Phys.Rev.D 65 (2002) 125019;\par 
          A. DiPiazza, G.Calucci, Phys.Rev.D 66 (2002) 123006;\par
          A. DiPiazza, Eur. Phys. J.C 36 (2004) 25
\item {5.}T. Piran, Phys.Rep.314 (1999) 375;\par
          J. van Paradijs, C. Kouveliotou, R.A.M.J. Wijers, Annu.Rev. Astron.Astrophys. 38 (2000) 279:\par
          P. M\'esz\'aros, Annu.Rev.Astron.Astrophys. 40 (2002)137;\par
          T. Piran, Rev.Mod.Phys. 76 (2004) 1143 \par
          R. Ruffini, Nuovo Cimento B 119 (2004) 785
\item {6.}R.D.Peccei, H.R.Quinn, Phys.Rev.Lett. 38 (1977) 1440;\par
          R.D.Peccei, H.R.Quinn, Phys.Rev.D 16 (1977) 1791;\par
          F. Wilczek, Phys.Rev.Lett. 40 (1978) 279;\par
          S. Weinberg, Phys.Rev.Lett. 40 (1978) 223
\item {7.}L. Maiani, R.Petronzio, E.Zavattini Phys.Lett.B 175 (1986) 359
\item {8.}E. Zavattini, G. Zavattini, G. Ruoso, E. Polacco, E. Milotti, M. Karuza, U. Gastaldi, G. DiDomenico, F. DellaValle, R. Cimino, S. Carusotto, G.Cantatore, M. Bregant Phys.Rev.Lett. 96 (2006) 110406\par
          S. Lamoreaux Nature 41 (2006) 31
\item {9.} The CERN Axion solar telescope (C.E.Asleth {\it et al.}) Nucl.Phys.B (proc.suppl.) 110 (2002) 85
\item {10.} A.B. Migdal, V. Krainov {\it Approximation methods in quantum mechanics (Ch. 2)} (Benjamin, NewYork 1969)\par
           D.R.Bates ed. {\it Quantum theory, vol 1 (Ch. 8)}Academic press, London 1961
\item {11.} A. DiPiazza, G.Calucci, Mod.Phys.Lett. A20 (2005), 117  	   
\item {12.} M.Abramowitz and I.A.Stegun, {\it Handbook of mathematical functions} (Dover, 1964).
\item {13.} S. Hannestad, A. Mirizzi,G. Raffelt, J. Cosm. Astrop. Phys. 07002 (2005)
\vfill\end